\documentclass{article}

\usepackage[english]{babel}
\usepackage{arxiv}

\usepackage[utf8]{inputenc} % allow utf-8 input
\usepackage[T1]{fontenc}    % use 8-bit T1 fonts
\usepackage{hyperref}       % hyperlinks
\usepackage{url}            % simple URL typesetting
\usepackage{booktabs}       % professional-quality tables
\usepackage{amsfonts}       % blackboard math symbols
\usepackage{nicefrac}       % compact symbols for 1/2, etc.
\usepackage{microtype}      % microtypography
\usepackage{lipsum}
\usepackage{amsmath}
\usepackage[title]{appendix}
\usepackage{mathtools}
\usepackage{filecontents}
\usepackage{graphicx}
\usepackage{multirow}
\usepackage{amssymb}

\usepackage{cleveref}
\usepackage{csquotes}

\title{A Rolling Optimized Nonlinear Grey Bernoulli Model RONGBM(1, 1) and application in predicting total COVID-19 infected cases}

\usepackage{biblatex}
\author{
  Hoang Anh NGO \thanks{Corresponding author, Tel: (+33) 7 61 86 80 94, Email address: \href{mailto:hoang-anh.ngo@polytechnique.edu}{hoang-anh.ngo@polytechnique.edu}} \\
  École Polytechnique\\
  Institut Polytechnique de Paris\\
  91120 Palaiseau, FRANCE \\
    \And
 Thai Nam HOANG \\
  Department of Mathematics and Computer Science \\
  Beloit College\\
  WI 53511, United States \\
}

\begin{document}
\maketitle

\begin{abstract}
The Nonlinear Grey Bernoulli Model NGBM(1, 1) is a recently developed grey model which has various applications in different fields, mainly due to its accuracy in handling small time-series datasets with nonlinear variations. In this paper, to fully improve the accuracy of this model, a novel model is proposed, namely Rolling Optimized Nonlinear Grey Bernoulli Model RONGBM(1, 1). This model combines the rolling mechanism with the simultaneous optimization of all model parameters (exponential, background value and initial condition). The accuracy of this new model has significantly been proven through forecasting Vietnam's GDP from 2013 to 2018, before it is applied to predict the total COVID-19 infected cases globally by day. 
\end{abstract}

% keywords , can be removed

\section{Introduction}
At the beginning of 2020, there happened to be an out break of pneumonia, started from an unknown aetiology in Wuhan City, Hubei Province, China. A novel strain of coronavirus (COVID-19) was found from a cluster of patients in January 2020 \cite{li2020nCoV}. Some scientists proposed a hypothesis that this strain of virus might have originated from bats that could be traced back to Huanan South China Seafood Market, a seafood market \cite{zhou2020nCoV}. There has been evidence about person-to-person transmission since the middle of December 2019 \cite{li2020nCoV}.

The number of cases has increased exponentially since then and authorities all over the world have to promulgate laws to alleviate the widespread of disease. Europe made the strict act as to close borders of all Schengen area \cite{EUborder}. At the end of the first quarter, USA surpassed China to become the world's most infected territory at more than $100,000$ confirmed cases \cite{USsurpassed100k}. Roughly a month later, USA, again, became the first country to surpass $1$ million confirm cases, however, the actually cases could be 10 times higher due to the lack of test kits \cite{USsurpassed1M,USlacktest}. 
Most lately, after half a year of outbreak, the total confirm cases all over the world has reached $10$ million, with more than $500,000$ people died. No country has succeeded in producing vaccination or special drugs, while the current situation is extremely complex and unpredictable, the need of data modelling and prediction is a must.

The number of COVID-19 infected cases can be regarded as a time series problem. According to previous research, time-varying process forecasting is one of the most important fields in terms of statistics. By collecting and analyzing different data points, researchers could investigate the situation and surmount current problem. However, most of these problems require a significant amount of data, which might be difficult due to the lack of information, or the ambiguity and non-linearity of data. For example, data of exchange rate between two currencies may fluctuate throughout a day or a period of time. Essentially, there is an urge to develop a method to solve this real-world dilemma. 

A lot of methodologies have been developed to solve the time series forecasting problems. Some can be implied, including traditional statistical methods like AR (Autoregressive Model) \cite{jie2010AR,chen1993AR,oskar2019AR}, MA (Moving Average) \cite{hansun2013}, ARIMA (Autoregressive Integrated Moving Average) \cite{meyler1998ARIMA,ayodele2016ARIMA} with abilities to predict linear trend, but are limited to fuzzy data. Others algorithms that can break through the cyclic of information and predict it can be listed out as LSTM (Long Short-term Memory) \cite{gers2000LSTM, LSTMog}, SVM (Support Vector Machine) \cite{thissen2003SVM}, and Sliding Window \cite{mozaffari2015SW}. Additionally, there are some hybrid models that can outperform others with just limits information and can improve forecasting results. 

To deal with the limitation of data, Grey method has been brought into consideration. Grey method was proposed by Deng Ju-Long in March, 1982 \cite{julong1982grey}. In Grey system theory \cite{liu2006grey, julong1982grey}, considered the degree of information, a white system is a system in which all information is known, while a black system is a system contains all unknown parts. Additionally, a system which consists of both known and unknown information is called a Grey system \cite{julong1982grey}. According to Guo, R. \cite{guo2005grey}, the Grey system outperforms other methods with just a limited number of discrete data, to achieve an insignificant margin of error of prediction versus real value. The system of total COVID-19 infected cases can be treated as a Grey system due to its constraint of data. 

Mathematically, the traditional Grey predicting model is based on the least square reduction and the first-order linear ordinary differential equation. For instance, traditional GM(1, 1) can be taken into account \cite{li2006gm, zhou2013gm,wang2009power} to solve this problem. Notably, to reinforce the predicting accuracy, some researchers have studied and combined to create hybrid Grey models, such as grey-Markov \cite{li2006gm,mao2011greymarkov}, grey-Fourier \cite{2014greyfourier}, grey-Taguchi \cite{equbal2019taguchi}. Some more hybrid models such as Grey rolling mechanism \cite{akay2007rolling,chang2005power,liu2014rolling} or nonlinear Grey Bernoulli method \cite{wang2011ongbm,chen2006NGM,wu2019chinagdp,lu2016optimize} can also be taken into consideration. In this paper, rolling mechanism is combined with nonlinear Grey Bernoulli method to propose a rolling optimized nonlinear Grey Bernoulli model, or RONGBM(1, 1). For benchmarking, AR, Artificial Neural Network (ANN), and LSTM are used to compare the accuracy with RONGBM(1, 1). This hybrid model is used to optimize the accuracy of predicting the total COVID-19 infected cases with just 11 entries of data (from January 28th, 2020 to February 7rd, 2020). The beginning date was chosen due to the fact that, there might have been uncleared data on those previous days.

The remainder of this paper will be organized as follows. In Section 2, we will provide a brief description of the traditional Grey models, including GM(1, 1) and the Nonlinear Grey Bernoulli Model GM(1, 1). Section 3 briefly introduces methods of optimizing parameters simultaneously in the formation of the optimized ONGBM(1, 1) model. In section 4, the application of the rolling mechanism is presented to propose a new RONGBM(1, 1) model. Section 5 provides some statistical and modern machine learning models as a benchmarking tool for the proposed model in the previous section. Section 6 introduces some modern machine learning models as an additional benchmarking tool for the newly proposed model in the empirical examples. Section 7 adopts the example of predicting Vietnam's GDP to demonstrate the accuracy of the new RONGBM(1, 1) model in comparison with models in Section 5, then applies this new model into predicting the total COVID-19 infected cases until \texttt{2020-02-18}. Section 8 concludes the paper by drawing some conclusions and future work.   

\section{Description of the traditional GM(1, 1) and NGBM(1, 1) models}
\label{sec:headings}

In this section, a brief introduction of the mathematical modelling of two traditional models: GM(1, 1) and Nonlinear Grey Bernoulli Model NGBM(1, 1) is introduced. 

\subsection{Building the traditional Grey model GM(1, 1)}
Assume that $x^{(0)}$ is the non - negative original historical time series of data with $m$ entries as:
    \begin{equation}
        x^{(0)} = \{ x^{(0)}(1), x^{(0)}(2), ..., x^{(0)}(k), ..., x^{(0)}(m)\}
    \end{equation}    
   
Next, we define $x^{(1)}$ using one-time accumulated generating operation (1 - AGO), which is as
   \begin{equation}
        x^{(1)} = \{ x^{(1)}(1), x^{(1)}(2), ..., x^{(1)}(k), ..., x^{(1)}(m)\}    
   \end{equation}
where
    \begin{equation}
        \begin{cases} 
        x^{(1)}(1) = x^{(0)}(1) \\
        x^{(1)}(k) = \displaystyle\sum_{i = 1}^k x^{(0)}(k), k = 2, 3, ..., m
        \end{cases}
    \end{equation}

As $x^{(1)}$ is a monotonic increasing sequence which is similar to the solution of a first-order linear differential equation, one can assume that the solution of the following differential equation
    \begin{equation}
        \frac{d\hat{x}^{(1)}}{dt} + a\hat{x}^{(1)} = b
    \end{equation}
represents the Grey predicted value complement to the initial condition $\hat{x}^{(1)}(1) = x^{(0)}(1)$ and parameters $a$ and $b$.

By definition, $\frac{d\hat{x}}{dt}$ can be written as:
    \begin{equation}
        \frac{d\hat{x}^{(1)}}{dt} = \lim_{\Delta t \rightarrow 0} \frac{\hat{x}^{(1)}(t + \Delta t) - \hat{x}^{(1)}(t)}{\Delta t}
    \end{equation}

However, to discretize the differential equation, $\Delta t$ can be set to be equal to $1$, which means that $(5)$ can be rewritten as 
    \begin{equation}
        \frac{d\hat{x}^{(1)}}{dt} = \hat{x}^{(1)}(t + 1) - \hat{x}^{(1)}(t) = x^{(1)}(t+1) - x^{(1)}(t) = x^{(0)}(t+1)
    \end{equation}

The Grey predicted value is now defined as
    \begin{equation}
        \hat{x}^{(1)}(t) \approx P x^{(1)}(k) + (1-P) x^{(1)}(k+1) = z^{(1)} (k+1), k = 1, 2, 3, ..., m
    \end{equation}
with $P$ usually set as $\frac{1}{2}$ in the traditional models.

The differential equation can now be discretized as:
    \begin{equation}
        x^{(0)}(k) + az^{(1)}(k) = b
    \end{equation}
    
By the least-squared method, the coefficients $a$ and $b$ can be determined by
    \begin{equation}
        \left[ \begin{matrix} a \\ b \end{matrix} \right] = \left( B^T B \right)^{-1} B^T Y
    \end{equation}
with 
    \begin{equation}
        B = \left[ \begin{matrix} - z^{(1)}(2) & 1 \\ - z^{(1)}(3) & 1 \\ ... \\ - z^{(1)}(m) & 1 \end{matrix} \right], Y = \left[ \begin{matrix} x^{(0)}(2) \\ x^{(0)}(3) \\ ... \\ x^{(0)}(m) \end{matrix} \right]
    \end{equation}
    
The particular solution of equation $(4)$ with the initial condition is:
    \begin{equation}
        \hat{x}^{(1)}(k+1) = \left( x^{(0)}(1) - \frac{b}{a} \right) e^{-ak} + \frac{b}{a}, k = 1, 2, 3, ..., m-1
    \end{equation}
    
The prediction of the historical time series of data at point $k+1$ can now be deduced by:
    \begin{equation}
        \hat{x}^{(0)}(k+1) = \hat{x}^{(1)}(k+1) - \hat{x}^{(1)}(k) = (1 - e^{-a}) \left( x^{(0)}(1) - \frac{b}{a} \right) e^{-ak}
    \end{equation}
with 
    \begin{equation}
        \begin{cases}
            \hat{x}^{(0)}(k) = x^{(0)}(k), k = 1, 2, ...,m \text{ fitted values, } \\
            \hat{x}^{(0)}(m+1), \hat{x}^{(0)}(m+2), ..., x^{(0)}(m+h) \text{ predicted values}
        \end{cases}
    \end{equation}

\subsection{The Nonlinear Grey Bernoulli Model NGBM(1, 1)}
In order to obtain higher accuracy in predicting comparing to the original GM(1, 1) model, Professor Chen \parencite{chen2008nbgm} proposed the Nonlinear Bernoulli Grey Model NBGM(1, 1) as follows:

Similarly to the traditional Grey model GM(1, 1), assume that $x^{(0)}$ is the non - negative original historical time series of data with $m$ entries as:
    \begin{equation}
        x^{(0)} = \{ x^{(0)}(1), x^{(0)}(2), ..., x^{(0)}(k), ..., x^{(0)}(m)\}
    \end{equation}    
   
Next, we define $x^{(1)}$ using one-time accumulated generating operation (1 - AGO), which is as
   \begin{equation}
        x^{(1)} = \{ x^{(1)}(1), x^{(1)}(2), ..., x^{(1)}(k), ..., x^{(1)}(m)\}    
   \end{equation}
where
    \begin{equation}
        \begin{cases}
        x^{(1)}(1) = x^{(0)}(1) \\
        x^{(1)}(k) = \displaystyle\sum_{i = 1}^k x^{(0)}(k), k = 2, 3, ..., m
        \end{cases}
    \end{equation}

As indicated previously, equation $(4)$ is a linear differential equation. A similar form of this equation, which is nonlinear and has the form of:
    \begin{equation}
        \frac{d\hat{x}^{(1)}}{dt} + a\hat{x}^{(1)} = b\left[\hat{x}^{(1)} \right]^n
    \end{equation}
where $n \in \mathbb{R}$ any real number is called a Bernoulli equation, or the with then differential equation of the NGBM(1, 1) model.

The background value is now also defined as:
    \begin{equation}
        z^{(1)}(k+1) = (1-P) x^{(1)}(k) + P x^{(1)}(k+1)
    \end{equation}
with $P = \frac{1}{2}$ for the traditional model. Then, discretizing the ODE, one obtains
    \begin{equation}
        x^{(0)}(k) + az^{(1)}(k) = b \left[ x^{(1)}(k) \right]^n
    \end{equation}
which is called the basic Grey differential equation of the NGBM(1, 1) model. It can easily be recognized that for $n = 0$, this equation turns to equation $(4)$, which is the traditional GM(1, 1) model; for $n = 2$, the equation turns to the Grey-Verhulst equation.

By the least square method, the parameters $a$ and $b$ can be determined by:
    \begin{equation}
        \left[ \begin{matrix} a \\ b \end{matrix} \right] = \left( B^T B \right)^{-1} B^T Y
    \end{equation}
with 
    \begin{equation}
        B = \left[ \begin{matrix} - z^{(1)}(2) & \left[ z^{(1)}(2)\right]^n \\ - z^{(1)}(3) & \left[ z^{(1)}(3)\right]^n \\ ... \\ - z^{(1)}(m) & \left[ z^{(1)}(m)\right]^n \end{matrix} \right], Y = \left[ \begin{matrix} x^{(0)}(2) \\ x^{(0)}(3) \\ ... \\ x^{(0)}(m) \end{matrix} \right]
    \end{equation}

The particular solution of equation $(17)$, or the discrete time function, with the initial condition is:
    \begin{equation}
        \hat{x}^{(1)}(k) = \left[ \left( x^{(0)}(1)^{(1-n)} - \frac{b}{a} \right) e^{-a(1-n)(k-1)} + \frac{b}{a} \right]^{\frac{1}{1-n}}, k = 1, 2, 3, ..., m
    \end{equation}
    
The prediction of the historical time series of data at point $k$ can now be deduced by:
    \begin{equation}
        \hat{x}^{(0)}(k) = \hat{x}^{(1)}(k) - \hat{x}^{(1)}(k-1)
    \end{equation}
    
\section{The Optimized Nonlinear Grey Bernoulli Model ONGBM(1, 1)}

\subsection{Optimization of the exponential parameter}

Based on the basic principle of the information overlapping of the Grey system, Wang et al. (2009) \parencite{wang2009power} proposed a formula determining the value of the exponential parameter by:
    \begin{equation}
        n = \frac{1}{m-2} \displaystyle\sum_{k = 2}^{m-1} \gamma(k)
    \end{equation}
with 
    \begin{equation}
        \gamma(k) = \frac{\splitfrac{\left[ x^{(0)}(k+1) - x^{(0)}(k) \right] \times z^{(1)}(k+1) \times z^{(1)}(k) \times x^{(0)}(k)}{- \left[ x^{(0)}(k) - x^{(0)}(k-1)\right] \times z^{(1)}(k+1) \times z^{(1)}(k) \times x^{(0)}(k+1)}}{\left[ x^{(0)}(k+1) \right]^2 \times z^{(1)}(k) \times x^{(0)}(k) - \left[ x^{(0)}(k) \right]^2 \times z^{(1)}(k+1) \times x^{(0)}(k+1)}
    \end{equation}
    
However, with the existence of modern programming languages, this formula yields the problem of inflexibility. Moreover, as further investigated in \textbf{Section 7}, this formula yield a result which is significantly less effective, comparing to the method of iterating on a pre-defined range of the exponential parameter of $[-1,1)$. 

\subsection{Optimization of the background value estimation}

According to Wang et al. (2011) \parencite{wang2011ongbm}, the precision of any Grey model is highly dependent on the value of $P$ in the background value $z^{(1)}(k)$. Change et al. (2005) \parencite{chang2005power} produced an even stronger statement by stating that forecasting accuracy of a Grey prediction model can be improved by optimizing parameter $P$. 

The background value $z^{(1)}(k)$ of any (derived) Grey model is considered to be an approximation of the integral region generated by the curve of the function $x^{(1)}$ and the abscissa axis within the interval $[k-1,k]$. Mathematically, this value is expressed as:
    \begin{equation}
        z^{(1)}(k) = \int_{k-1}^k x^{(1)} dx
    \end{equation}
Applying the mean value theorem, it can be derived that:
    \begin{equation}
        z^{(1)}(k) = P x^{(1)}(k) + (1-P) x^{(1)}(k-1), P \in [0,1]
    \end{equation}
which are exactly equations $(8)$ and $(18)$. Mentioned beforehand, $P = \frac{1}{2}$ is the usually chosen values for traditional Grey models. However, it is trivial that the accuracy of the model depends heavily on the parameter $P$, which means that fixing this parameter at $\frac{1}{2}$ as usual will reduce massively the effectiveness when there are values fluctuating and deviating from the overall trend. 

Zhang (1993) \parencite{zhuang1993para} has proven that the parameter $P$ determining the background values and parameter $a$ satisfy the following condition:
    \begin{equation}
        P = \frac{1}{a} - \frac{1}{e^a - 1}
    \end{equation}

From this formula, we derive:
    \begin{equation}
        \begin{split}
            \lim_{a \to 0} P & = \lim_{a \to 0} \frac{1}{a} - \frac{1}{e^a - 1} = \lim_{a \to 0} \frac{e^a - 1 - a}{a(e^a - 1)} \\ 
            & = \lim_{a \to 0} \frac{e^a - 1}{ae^a + e^a - 1} \left( \text{L'Hôpital rule on } \frac{f(x)}{g(x)} \right) \\
            & = \lim_{a \to 0} \frac{e^a}{ae^a + e^a + e^a} \left( \text{L'Hôpital rule on } \frac{f'(x)}{g'(x)} \right) \\
            & = \frac{e^0}{0 \times e^0 + e^0 + e^0}  \\ 
            & = \frac{1}{2}
        \end{split}
    \end{equation}

which is a valid explanation for the setting of the value of $P$ to $\frac{1}{2}$ for traditional Grey prediction models. However, when the value of $a$ is significant (the absolute value of $a$ is large enough), customarily setting the coefficient of $P$ to $\frac{1}{2}$ is incorrect, reducing the efficiency of the model. 

As a result, a new calculating formula was proposed by Tan (2000) \parencite{tan2000background} as:
    \begin{equation}
        z^{(1)}(k) = \frac{1}{2q} \left[ (q+1) x^{(1)}(k-1) + (q-1) x^{(1)}(k) \right], k = 2, 3, ..., m
    \end{equation}

with $q$ being given empirically by \parencite{lu2016optimize}:
    \begin{equation}
        q = \left( \displaystyle\sum_{i = 2}^n \frac{x^{(1)}(k)}{x^{(1)}(k-1)} \right)^{\frac{1}{m-1}} + (m-1) 
    \end{equation}
    
This means that, we have defined the value of $P$ by:
    \begin{equation}
        P = \frac{q+1}{2q} = \frac{1}{2} + \frac{1}{2q}
    \end{equation}
    
Once again, this formula yields the problem of inflexibility. As 1-AGO sequence is a strictly increasing sequence, 
    \begin{equation}
        \frac{x^{(1)}(k)}{x^{(1)}(k-1)} > 0, k = 2, 3, ..., m
    \end{equation}
which means that 
    \begin{equation}
        q = \left( \displaystyle\sum_{i = 2}^m \frac{x^{(1)}(k)}{x^{(1)}(k-1)} \right)^{\frac{1}{m-1}} + (m-1) > 0
    \end{equation}
leading to
    \begin{equation}
        P = \frac{1}{2} + \frac{1}{2q} > \frac{1}{2}
    \end{equation}
which has already omitted the range of $\left[ 0, \frac{1}{2} \right]$ for consideration. Moreover, as applied later, upon applying this formula, the algorithm yields higher error than the empirically derived value of the same parameter. The optimal solution for this problem would also be iterating on a well-defined range of values of $P$ ($[0,1]$) with a small enough step size.

Within the previous session, the parameters $P$ and $n$ are not optimized simultaneously. 

\subsection{Optimization of the initial condition}

Considering the particular solution of equation $(17)$, the initial condition is set to be $x^{(0)}(1)$, the first point of the original time series of data. Using the principle of new information prior choosing, the last item of the 1 - AGO series, $x^{(1)}(m)$, can be used to enhance the accuracy of the model. Dang (2004) \parencite{dang2004xn} has confirmed this theory by testing on two different Grey models: GM(1, 1) and the Grey-Verhulst model. Replacing $x^{(1)}(m)$ for $^{(0)}(1)$ as the initial condition, the time response function for the whitening equation of the NGBM(1, 1) model can be rewritten as:
        \begin{equation}
            \hat{x}^{(1)}(k) = \left[ \left( x^{(1)}(m)^{(1-n)} - \frac{b}{a} \right) e^{-a(1-n)(k-m)} + \frac{b}{a} \right]^{\frac{1}{1-n}}
        \end{equation}

Furthermore, since the model parameters $a$ and $b$ are obtained with least square method, there will be high chances that the model will not pass through the point $x^{(0)}(n)$, As a result, the new initial condition $x^{(1)}(m)$ will require correction in order to maximize optimization. This means that the new initial term, instead, would be $x^{(1)}(m) + c$.

The correction parameter $c$ would be chosen in order to minimize the following function ( proposed by Lu et al. 2016 \parencite{lu2016optimize}): 
    \begin{equation}
        f(c) = \displaystyle\sum_{k = 1}^m \left\{ \left[ \hat{x}^{(1)}(k) \right]^{1 - n} - \left[ x^{(1)}(k)\right]^{1-n} \right\}^2
    \end{equation}

Substituting the discrete time response function deduced from \textbf{Theorem 1}, we obtain:
    \begin{equation}
        f(c) = \displaystyle\sum_{k = 1}^m \left\{ \frac{b}{a} + \left[ \left( x^{(1)}(m) + c \right)^{1 - n} - \frac{b}{a} \right] e^{-a(1-n)(k-m)} - \left[ x^{(1)}(k)\right]^{1-n} \right\}^2
    \end{equation}
    
Let 
    \begin{equation}
        \begin{cases}
            E(k) = e^{-a(1-n)(k-m)} \\
            A(k) = \left[ x^{(1)}(k) \right]^{1 - n} - \frac{b}{a} \left( 1 - E(k) \right) 
        \end{cases}
    \end{equation}
Then, $f(c)$ will be rewritten as:
    \begin{equation}
        f(c) = \displaystyle\sum_{k = 1}^m \left\{ \left[ x^{(1)}(k) + c \right]^{1 - n} E(k) - A(k) \right\}^2
    \end{equation}
    
To find the minimum value of $f(c)$, we set the first - order condition (FOC) to $0$, which means that $\frac{\partial f(c)}{\partial c} = 0$, or:
    \begin{equation}
        2(1-n)\left[ x^{(1)}(k) + c \right]^{-n} \displaystyle\sum_{k = 1}^m \left\{ \left[ x^{(1)}(k) + c\right]^{1-n} E(k)^2 - A(k) E(k) \right\} = 0
    \end{equation}

We deduce
    \begin{equation}
        \left[ x^{(1)}(m) + c\right]^{1-n} = \frac{\displaystyle\sum_{k = 1}^m A(k) E(k)}{\displaystyle\sum_{k = 1}^m E(k)^2}
    \end{equation}

which means that the discrete time response function of the NGBM(1, 1) model can be now rewritten as:
    \begin{equation}
        \hat{x}^{(1)}(k) = \left[ \left( \frac{\displaystyle\sum_{k = 1}^m A(k) E(k)}{\displaystyle\sum_{k = 1}^m E(k)^2} - \frac{b}{a}\right) e^{-a(1-n)(k-m)} + \frac{b}{a} \right]^{\frac{1}{1-n}}
    \end{equation}

\subsection{Proposing algorithm}

Based on the simultaneous optimization of three parameters: the exponential parameter, the initial condition and the background value, we propose a further optimized algorithm of the Optimized NGBM(1, 1) given by Wu et al. (2019) \parencite{wu2019chinagdp}. This algorithm is more practical and less time-consuming upon implementing in the currently available computer languages (R, Python or Matlab):
    \begin{enumerate}
        \item Given the original time series of the data $x^{(0)} = \{ x^{(0)}(1), x^{(0)}(2), ..., x^{(0)}(m)\}$, calculate the 1 - AGO sequence $x^{(1)}$.
        \item Indicate the range of values for $P$ and $n$. According to the mean value theorem, $P$ must belong to the range $[0,1]$, while $n$ is usually considered within the range of $[-1,1)$. Considering currently available computational tools, the step size is usually set at $0.001 - 0.01$. 
        
        \textbf{Remark:} Empirically, for $n \notin [-1,1)$, the model has been proven to yield higher error on average, reducing the efficiency of the model. As a result, the values of $n$ outside the interval is not within our interest.
        \item For the pair of values of $P$ and $n$, calculate the matrices $B$ and $Y$, which yields parameters $a$ and $b$. 
        \item Calculate RPE and ARPE for each $a$ and $b$. Then, choose the parameters which yield the minimum ARPE. 
        \item Calculate the values of $E(k)$ and $A(k)$ to deduce the error correction first initial condition $x^{(1)}(m) + c$. Then, compute $\hat{x}^{(1)}$ based on the proposed discrete time response function $(44)$.
        \item From $\hat{x}^{(1)}$, using the inverse AGO, compute $\hat{x}^{(0)}$.
    \end{enumerate}
    
\section{The Rolling Optimized Nonlinear Grey Bernoulli Model RONGBM(1, 1)}
As mentioned by Akay and Atak (2007) \parencite{akay2007rolling}, the rolling mechanism is an efficient technique to increase forecasting accuracy of Grey prediction when there are exponential increasing/decreasing or chaotic data. This mechanism provides the ability to update input data by removing the oldest data points for each loop before applying the prediction algorithm. Its purpose is that, for each rolling step, the set of data points used for the upcoming forecasting is the most recent.

Wu et al. (2019) \parencite{wu2019chinagdp} is the first paper to ever mention the application of rolling mechanism to the NGBM(1, 1) to enhance accuracy. However, the proposed model did not reach the maximum efficiency due to the following reasons:
    \begin{enumerate}
        \item In the proposed algorithm, the optimization of the parameters $P$ and $n$ are determined by running a loop through a set of pre-defined value by the user. In this case, the author defined the set of values of $P$ to be between $0$ and $1$, with the step of $0.01$; while $n$ is considered between $-1$ and $0.99$, with the step of $0.01$. This step size for both parameters is still large enough to miss the optimal value.
        \item In \textbf{Algorithm 1}, the background value is set as 
            \begin{equation}
                z^{(1)}(k) = P (x^{(1)}(k) + x^{(1)}(k+1)), k = 1, 2, ..., m
            \end{equation}
        which is not the optimal approximation and not compatible with the mean value theorem.
        \item As proposed, re-initiating the initial value to $x^{(1)}(m)$ will enhance accuracy of the algorithm. However, this can further be improved by correcting this initial value itself with the error correction parameter $c$ proposed in the previous section.
        \item The rolling mechanism implemented in \textbf{Algorithm 1} of \parencite{wu2019chinagdp} is incorrect. The rolling mechanism requires the insertion of the predicted value into the new array, while removing the oldest data point. In the algorithm, however, the author inserted the actual value instead of the predicted value, which develops biases for this model. 
    \end{enumerate}

Assume that $x^{(0)}$ is a non - negative input of time series of data with $m$ entries as:
    \begin{equation}
        x^{(0)} = \{ x^{(0)}(1), x^{(0)}(2), ..., x^{(0)}(k), ..., x^{(0)}(m)\}
    \end{equation}
and the output sequence of predicted data would be:
    \begin{equation}
        \hat{x}^{(0)} = \{ \hat{x}^{(0)}(1), \hat{x}^{(0)}(2), ..., \hat{x}^{(0)}(k), ..., \hat{x}^{(0)}(m), \hat{x}^{(0)}(m+1), ..., \hat{x}^{(0)}(M) \}, M \geq m+1, M \in \mathbb{N}
    \end{equation}

We propose a rolling Optimized NGBM(1, 1) model with the following algorithm:
    \begin{enumerate}
        \item Set $i = 1$.
        \item Set the training data to $x^{(0)} = \{x^{(0)}(1), x^{(0)}(2), ..., x^{(0)}(m) \}$. Use this training data to predict the following data points:
            \begin{equation}
                \{x^{(0)}(1), x^{(0)}(2), ..., x^{(0)}(m), x^{(0)}(m+1) \}
            \end{equation}
        \item While $i \leq M - m$:
            \begin{enumerate}
                \item Set the new value of $i$ as $i := i + 1$
                \item Set the training data to $x^{(0)} = \{x^{(0)}(i), x^{(0)}(i+1), ..., x^{(0)}(m + i -1) \}$.
                \item Apply the Optimized NGBM(1, 1) model on this new training data to predict $\hat{x}^{(0)}(m+i)$. 
                
                \textbf{Remark:} For each $i$, there will be different values of the set of the model parameters $\{a, b, P, n\}$.
                \item Return to the beginning of this step.
            \end{enumerate}
        \item When $i = M - m$, we have had all the predicted values $\{ \hat{x}^{(0)}(m+1), \hat{x}^{(0)}(m+1), ..., \hat{x}^{(0)}(M)\}$. 
        \item Our output sequence of predicted data would now be:
            \begin{equation}
                \hat{x}^{(0)} = \{ x^{(0)}(1), x^{(0)}(2), ..., x^{(0)}(k), ..., x^{(0)}(m), \hat{x}^{(0)}(m+1), ..., \hat{x}^{(0)}(M) \}
            \end{equation}
    \end{enumerate}

\section{Performance evaluation}
In order to examine the precision of each model, it is necessary to evaluate the difference between the fitted values and the actual values. For each specific time stamp $k$, the relative percentage error (RPE) calculates the relative difference of the forecast and recorded value by:
    \begin{equation}
       RPE = \epsilon(k) = \frac{x^{(0)}(k) - \hat{x}^{(0)}(k)}{x^{(0)}(k)} \times 100\%, k = 1, 2, 3, ..., m 
    \end{equation}
where $x^{(0)}(k)$ is the actual value and $\hat{x}({(0)}(k)$ is the fitted value.

The overall precision of the model can be measured by the average relative percentage error (ARPE), which is determined by:
    \begin{equation}
        ARPE = \epsilon (avg) = \displaystyle\sum_{k = 1}^m |\epsilon(k)| = \displaystyle\sum_{k = 1}^m \left| \frac{x^{(0)}(k) - \hat{x}^{(0)}(k)}{x^{(0)}(k)}\right| \times 100\% 
    \end{equation}

The classification of the model precision based on ARPE is described in Table ~\ref{table:ARPEclass} below \parencite{wu2019chinagdp}.

\begin{table}[h]
    \caption{ARPE classification of model precision}
    \label{table:ARPEclass}
    \centering
    \begin{tabular}{lllll}
        \toprule
        ARPE ($\%$)     & $\leq 10$     & $10 \sim 20$ & $20 \sim 50$ & $\geq 50$ \\
        \hline
        Classification & Excellent  & Good & Reasonable & Unacceptable     \\
        \bottomrule
  \end{tabular}
  \label{tab:table}
\end{table}

Besides ARPE, the root-mean-square error (RMSE) or root-mean-square deviation (RMSD) is also used to evaluate the precision of the model:
    \begin{equation}
        RMSE = \sqrt{\frac{\displaystyle\sum_{k = 1}^m \left(x^{(0)}(k) - \hat{x}^{(0)}(k) \right)^2}{m}}
    \end{equation}

Based on the concept of RMSE, the posterior error ratio ($c$) is established and is also to evaluate the efficiency of a model:
    \begin{equation}
        c = \frac{\sqrt{\frac{1}{n} \displaystyle\sum_{k = 1}^m (\epsilon(k) - \epsilon(avg))^2}}{\sqrt{\frac{1}{n} \displaystyle\sum_{k = 1}^m (x^{(0)}(k) - \bar{x}^{(0)})^2}}
    \end{equation}
    
The precision classification with respect to different levels of posterior error ratio is described in the following table ~\ref{table:posteriorratio} \parencite{chang2013media}.

\begin{table}[h]
    \caption{Posterior error ratio and precision rank}
    \centering
    \label{table:posteriorratio}
    \begin{tabular}{lllll}
        \toprule
        Posterior error ratio ($c$)     & $\leq 0.35$     & $0.35 \sim 0.5$ & $0.5 \sim 0.65$ & $\geq 0.65$ \\
        \hline
        Classification & 1 (Highly accurate)  & 2 (Qualified) & 3 (Marginal) & 4 (Disqualified)     \\
        \bottomrule
  \end{tabular}
\end{table}

Within the scope of this paper, RPE and ARPE will be used as the main tool of benchmarking and comparing efficiency between different models. This approach is the most popular and widely-used upon considering empirical results of any newly-proposed Grey model.
\section{Comparison with alternative algorithms}
In this section, we will indicate some alternative machine learning algorithms based on neural network. For this paper, we used 2 Neural Network models: Artificial Neural Network (ANN) and Long Short-term Memory (LSTM). Historically, recurrent Neural Network (RNN) and Convolutional Neural Network (CNN) were first introduced to handle such heavy-load tasks, like Image Processing or Natural Language Processing, and thus they were too complex for typical time series problems. However, researchers have started to adopt these deep learning methods, and turned them into perfect fits for a time-varying process. There are 2 main properties that make NN a competitive candidate for time series. NNs have general nonlinear function mapping capability that can approximate any continuous function. This gives NNs a capability of handling any complex problems with limited data. Moreover, NN is a non-parametric data-driven model, hence it does not require (mostly) any constraints. This is the key point for solving nonlinear, non-pattern problems that may have unique characteristics that cannot be captured by parametric model \cite{oskar2019AR}. 

\subsection{Artificial Neural Network (ANN)}
Artificial Neural Network (ANN) is a brain-inspired system, which simulates the way human learns. A single layer neural network is called a Perceptron, which can give a single output. 

    \begin{figure}[h]
      \centering
      \includegraphics[scale=0.5]{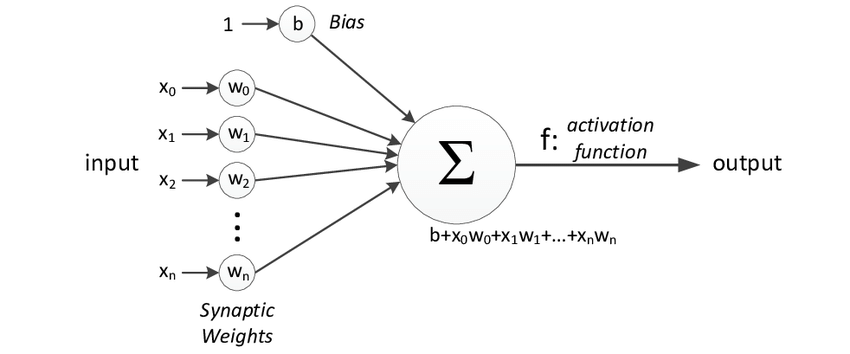}
      \caption{A Perception \cite{TDS_ANN}}
      \label{fig:fig1}
    \end{figure}

To use ANN, we have to decide an Activation function, which converts an input signal of a node in an ANN to an output signal. This output signal is used as input to the next layer in the stack \cite{TDS_ANN}. There are 4 Activation functions that can be taken into consideration. 

    \begin{itemize}
        \item Threshold Activation function (Binary Step function)
        A threshold-based binary activation function. For classification only (because of activated cases only 1 or 0).
            \begin{equation}
            f(x) = 
                \begin{cases}
                0, x < 0 \\
                1, x \geqslant 0
                \end{cases}
            \end{equation}
            
            \item Sigmoid Activation function (Logistic function)
            A S-shaped curve or sigmoid, ranges between 0 and 1. For predicting the probability
                \begin{equation}
                    \phi (z) = \frac{1}{1 + e^{-z}}
                \end{equation}
                
            \item Hyperbolic Tangent function (tanh)
            A similarity to sigmoid, but better in performance, ranges between -1 and 1. Perfect fit for layer stacking.
                \begin{equation}
                    f(x) = tanh(x)
                \end{equation}
                
            \item Rectified Linear Units (ReLU)
            The most common Activation function in both CNN and ANN, ranges from zero to infinity $[0, \infty)$.
    \end{itemize}
    
Based on the difference between the actual value and the predicted value, an error value called Cost Function is computed and sent back through the system. Between an input layer and an output layer is at least a hidden layer. The hidden layers perform computations on the weighted inputs and produce net input which is then applied with activation functions to produce the actual output. ANN uses a procedure called Back-propagation cyclically until the error value is kept at minimum \cite{TDS_ANN}.  

    \begin{figure}[h]
      \centering
      \includegraphics[scale=0.35]{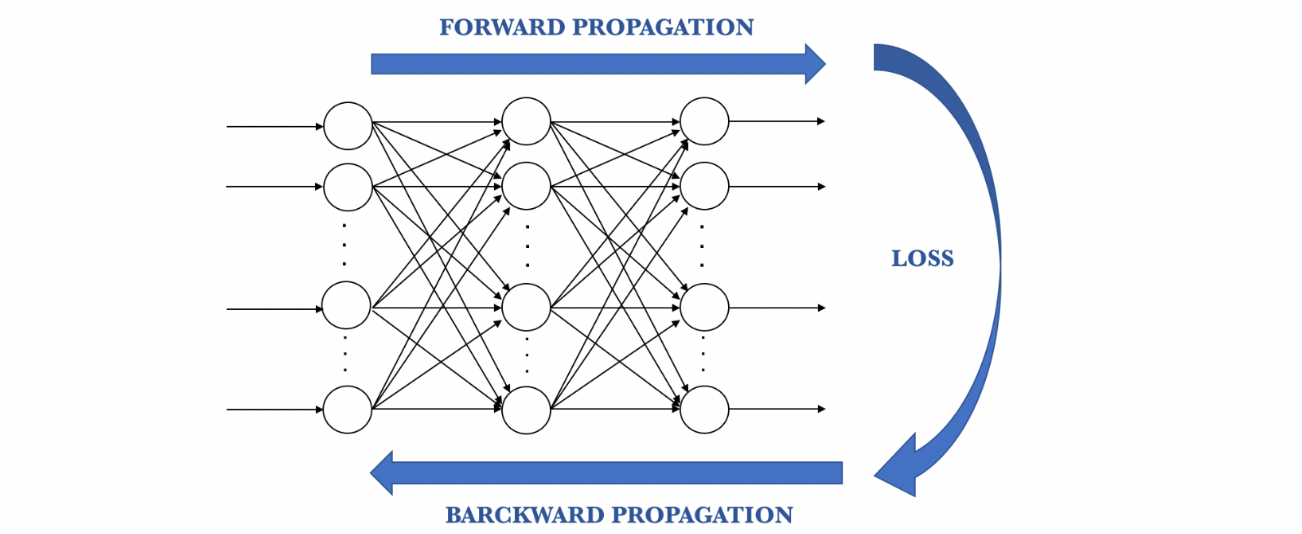}
      \caption{Back-Propagation procedure \cite{TDS_ANN}}
      \label{fig:fig2}
    \end{figure}

\subsection{Long Short-term Memory (LSTM)}
Long Short-term Memory (LSTM) is a RNN-based algorithm with the capability of learning long-term dependencies, which was proposed by Sepp Hochreiter and Jürgen Schmidhuber in  1997. Vanilla RNN method suffers from short-term memory during feed-forward. This means that, if processing data which has long time piece, it will be hard for RNN to carry (or remember) data from earlier time steps to later ones. Thus, RNN may leave out important information from the beginning. It also suffers from vanishing gradient problem in back-propagation. If a gradient value is too smal, it does not contribute much in learning process. 

LSTM was created as a solution to this problem. LSTM has feedback connection, which allows it to process the whole data sequence. A general LSTM consists of multiple cells, with forget gates, an input and an output. They have the ability to choose which information to carry on or leave behind. \cite{LSTMog,gers2000LSTM,colahLSTM}. 

    \begin{figure}[h]
        \centering
        \includegraphics{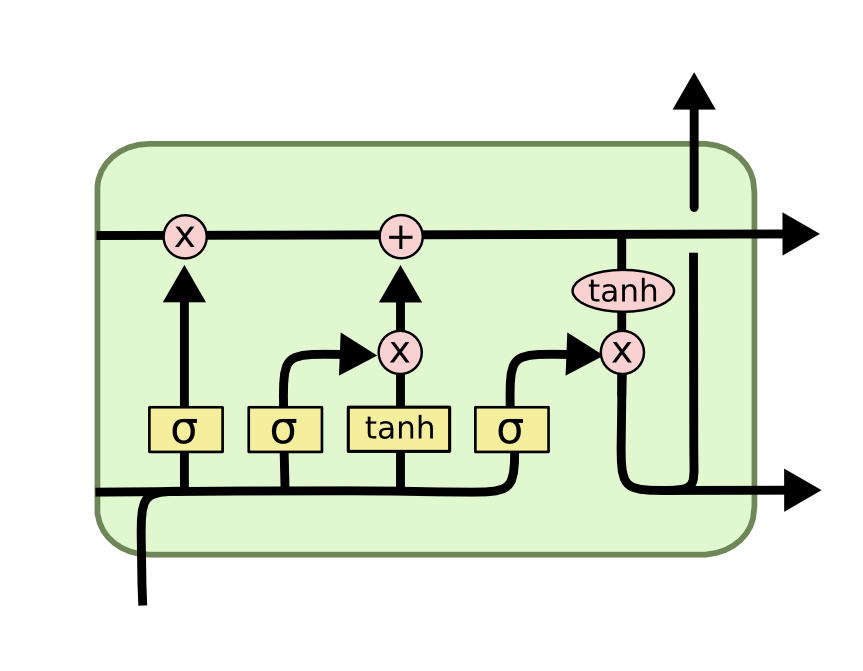}
        \caption{A LSTM cell \cite{colahLSTM}}
        \label{fig:fig3}
    \end{figure}

\section{Empirical applications}

\subsection{Efficiency testing for the Rolling Optmizied Nonlinear Grey Bernoulli Model RONGBM(1, 1)}

\subsubsection{Context}
After the end of the Vietnam War in Spring 1975, Vietnam's economic status started to slow down and decline. Furthermore, the embargo of the United States on Vietnam and the collapse of the Soviet Union led to the plummet of the Vietnam economy, hit rock-bottom in 1989 - 1990 \cite{VNgdp}. Since then, the economic status has revived significantly, with the annual growth rate of 6-8\% \cite{VNannualrate}. 

As mentioned above, we used GDP data between 2005 and 2018, as from 2005 and after, Vietnam's GDP has some certain growth in economic status and annual growth rate. Moreover, Vietnam is on the verge of development and integration into the world, using this range of data will enhance the forecasting accuracy, as it can avoid the inaccurate that those previous world events bring.

In this section, we will regard the most important macroeconomics variable as our testing data: the annual Gross Domestic Product (GDP) of Vietnam, between $2005$ and $2018$. These data can be retrieved from World Bank Open Data (\url{https://data.worldbank.org/})

\subsubsection{Methodology}

In order to demonstrate the accuracy and effectiveness of the newly proposed Rolling Optimized NGBM(1, 1) (abberivated as RONGBM(1, 1)), we will proceed to compare this model with the existing Grey models, including:
    \begin{itemize}
        \item The traditional Grey model GM(1, 1)
        \item The traditional Nonlinear Grey Bernoulli Grey Model NGBM(1, 1)
        \item The Optimized Nonlinear Grey Bernoulli Model ONGBM(1, 1). This model has adapted all possible optimization mentioned previously, including optimization of the background value, initial condition and exponential parameter.
    \end{itemize}
    
The data from $2004$ to $2013$ will be used for training, and data from $2014$ to $2018$ will be used for forecasting and testing models.

\subsubsection{Results and analysis}

From Table ~\ref{table:VNGDP05183models} and Table ~\ref{table:VNGDP0518RONGBM}, it can easily be observed that the ARPE generated by the newly proposed Rolling Optimized NGBM(1, 1) model is significantly smaller than that of comparing models. This is due to the fact that the initial condition $x^{(1)}(m) + c$ and the original training data have been updated, which yields closer result to the actual data. 

This empirical result also explains the conclusion driven in the previous sections that the estimating formula for the exponential parameter $n$ and the background value $P$ is not optimal, namely $(25)$ and $(31)$. Upon using the equations above, the values of $P$ and $n$ would be:
    \begin{equation}
        \begin{cases}
        P = 0.5484042 \\ n = -0.1387509
        \end{cases}
    \end{equation}
which yield higher ARPE ($2.8314\%$) for the first $10$ GDP values between $2004 - 2013$ comparing to the parameters derive empirically below with ARPE of $1.3855\%$.

Moreover, as mentioned above, the empirical value of $P = 0.495$ cannot be captured by equation $(31)$.
\begin{table}
    \begin{center}
    
    \caption{Results of GM(1, 1), NGBM(1, 1) and ONGBM(1, 1) in predicting Vietnam's GDP $2004 - 2018$}
    
    \label{table:VNGDP05183models}
    
    \begin{tabular}{|c|c|cc|cc|cc|}
    \hline
\multirow{5}{4em}{\centering Year} & \multirow{5}{6em}{\centering Actual Data} &  \multicolumn{2}{c}{\textbf{ }} & \multicolumn{2}{|c|}{Formula-optimized} & \multicolumn{2}{|c|}{Nash}\\ 

 &  &  &  & \multicolumn{2}{c}{NGBM(1, 1)} & \multicolumn{2}{|c|}{NGBM(1, 1)} \\ 

 &  &  \multicolumn{2}{c}{GM(1, 1)} & \multicolumn{2}{|c|}{$n=0.126$} & \multicolumn{2}{|c|}{$n=0.13,P=0.495$} \\ 
 &  &  &  &  &  &  & \\ 
 &  & \multirow{2}{5em}{Prediction} & RPE & \multirow{2}{5em}{Prediction} & RPE & \multirow{2}{5em}{Prediction} & RPE \\

 &  &  & (\%) &  & (\%) &  & (\%) \\ 
 
 \hline\hline

 \texttt{2004} & $45.42785$ & $45.42785$ & $0.00$ & $45.42785$ & $0.00$ & $45.42785$ & $0.00$\\ 
 \texttt{2005} & $57.63326$ & $61.43522$ & $6.60$ & $57.62228$ & $-0.02$ & $57.55257$ & $-0.14$\\
 \texttt{2006} & $66.37166$ & $70.01275$ & $5.48$ & $68.73623$ & $3.56$ & $68.75453$ & $3.59$\\
 \texttt{2007} & $77.41443$ & $79.78786$ & $3.07$ & $79.99618$ & $3.33$ & $80.07765$ & $3.44$\\
 \texttt{2008} & $99.13030$ & $90.92776$ & $-8.27$ & $91.99635$ & $-7.20$ & $92.12421$ & $-7.06$\\
 \texttt{2009} & $106.01466$ & $103.62301$ & $-2.26$ & $105.05247$ & $-0.91$ & $105.21288$ & $-0.75$\\
 \texttt{2010} & $115.93175$ & $118.09075$ & $1.86$ & $119.40416$ & $3.00$ & $119.58390$ & $3.15$\\
 \texttt{2011} & $135.53944$ & $134.57846$ & $-0.71$ & $135.27036$ & $-0.20$ & $135.45557$ & $-0.06$\\
 \texttt{2012} & $155.82000$ & $153.36817$ & $-1.57$ & $152.87099$ & $-1.89$ & $153.04632$ & $-1.78$\\
 \texttt{2013} & $171.22203$ & $174.78129$ & $2.08$ & $172.43780$ & $0.71$ & $172.58566$ & $0.79$\\
 \texttt{2014} & $186.20465$ & $199.18408$ & $6.97$ & $194.22121$ & $4.30$ & $194.32111$ & $4.35$\\
 \texttt{2015} & $193.24111$ & $226.99396$ & $17.47$ & $218.49546$ & $13.07$ & $218.52332$ & $13.08$\\
 \texttt{2016} & $205.27617$ & $258.68664$ & $26.02$ & $245.56317$ & $19.63$ & $245.49057$ & $19.59$\\
 \texttt{2017} & $223.77987$ & $294.80421$ & $31.74$ & $275.75975$ & $23.23$ & $275.55313$ & $23.13$\\
 \texttt{2018} & $245.21369$ & $335.96448$ & $37.01$ & $309.45795$ & $26.20$ & $309.07767$ & $26.04$\\
 
\hline\hline

ARPE &  &  & $\textbf{10.07}$ &  & $\textbf{7.15}$ &  & $\textbf{7.13}$ \\
Classification & & & Good & & Excellent & & Excellent \\
\hline
\end{tabular} 
\end{center}
\end{table}

\begin{table}
    \begin{center}
    \caption{Parameters and results of RONGBM(1, 1) in predicting Vietnam's GDP $2004 - 2018$}
    \label{table:VNGDP0518RONGBM}
    \begin{tabular}{|c|c|cc|cc|}
    \hline
\multirow{4}{4em}{\centering Year} & \multirow{4}{6em}{\centering Actual Data} &  \multicolumn{4}{c|}{Rolling Optimized} \\ 
 &  &  \multicolumn{4}{c|}{NGBM(1, 1)} \\ 
 &  &  &  &  & \\ 
 &  &  \multirow{2}{5em}{Parameter $P$} & \multirow{2}{5em}{Parameter $n$} & \multirow{2}{5em}{Prediction} & RPE \\ 
 &  &  &  &  &  (\%) \\ 
 
 \hline\hline
 \texttt{2004} & $45.42785$ & $0.495$ & $0.13$ & $45.42785$ & $0.00$\\ 
 \texttt{2005} & $57.63326$ & $0.495$ & $0.13$ & $57.55257$ & $-0.14$\\
 \texttt{2006} & $66.37166$ & $0.495$ & $0.13$ & $68.75453$ & $3.59$\\
 \texttt{2007} & $77.41443$ & $0.495$ & $0.13$ & $80.07765$ & $3.44$\\
 \texttt{2008} & $99.13030$ & $0.495$ & $0.13$  & $92.12421$ & $-7.06$\\
 \texttt{2009} & $106.01466$ & $0.495$ & $0.13$  & $105.21288$ & $-0.75$\\
 \texttt{2010} & $115.93175$ & $0.495$ & $0.13$ & $119.58390$ & $3.15$\\
 \texttt{2011} & $135.53944$ & $0.495$ & $0.13$ & $135.45557$ & $-0.06$\\
 \texttt{2012} & $155.82000$ & $0.495$ & $0.13$ & $153.04632$ & $-1.78$\\
 \texttt{2013} & $171.22203$ & $0.495$ & $0.13$ & $172.58566$ & $0.79$\\
 \texttt{2014} & $186.20465$ & $0.495$ & $0.13$ & $194.32111$ & $4.35$\\
 \texttt{2015} & $193.24111$ & $0.525$ & $0.165$ & $214.67979$ & $11.09$\\
 \texttt{2016} & $205.27617$ & $0.48$ & $0.155$ & $238.67530$ & $16.27$\\
 \texttt{2017} & $223.77987$ & $0.495$ & $-0.02$ & $270.80266$ & $21.01$\\
 \texttt{2018} & $245.21369$ & $0.47$ & $0.03$ & $303.5345$ & $23.78$\\
\hline\hline
  ARPE &  &  &  &  & $\textbf{6.48}$\\
  Classification & & & & & Excellent\\
\hline
\end{tabular} 
\end{center}
\end{table}
 
\begin{figure}[h]
    \centering
    \caption{Vietnam's GDP prediction ($2004 - 2018$) with GM(1, 1), NGBM(1, 1), ONGBM(1, 1), and RONGBM(1, 1)}
    \includegraphics[scale=0.3]{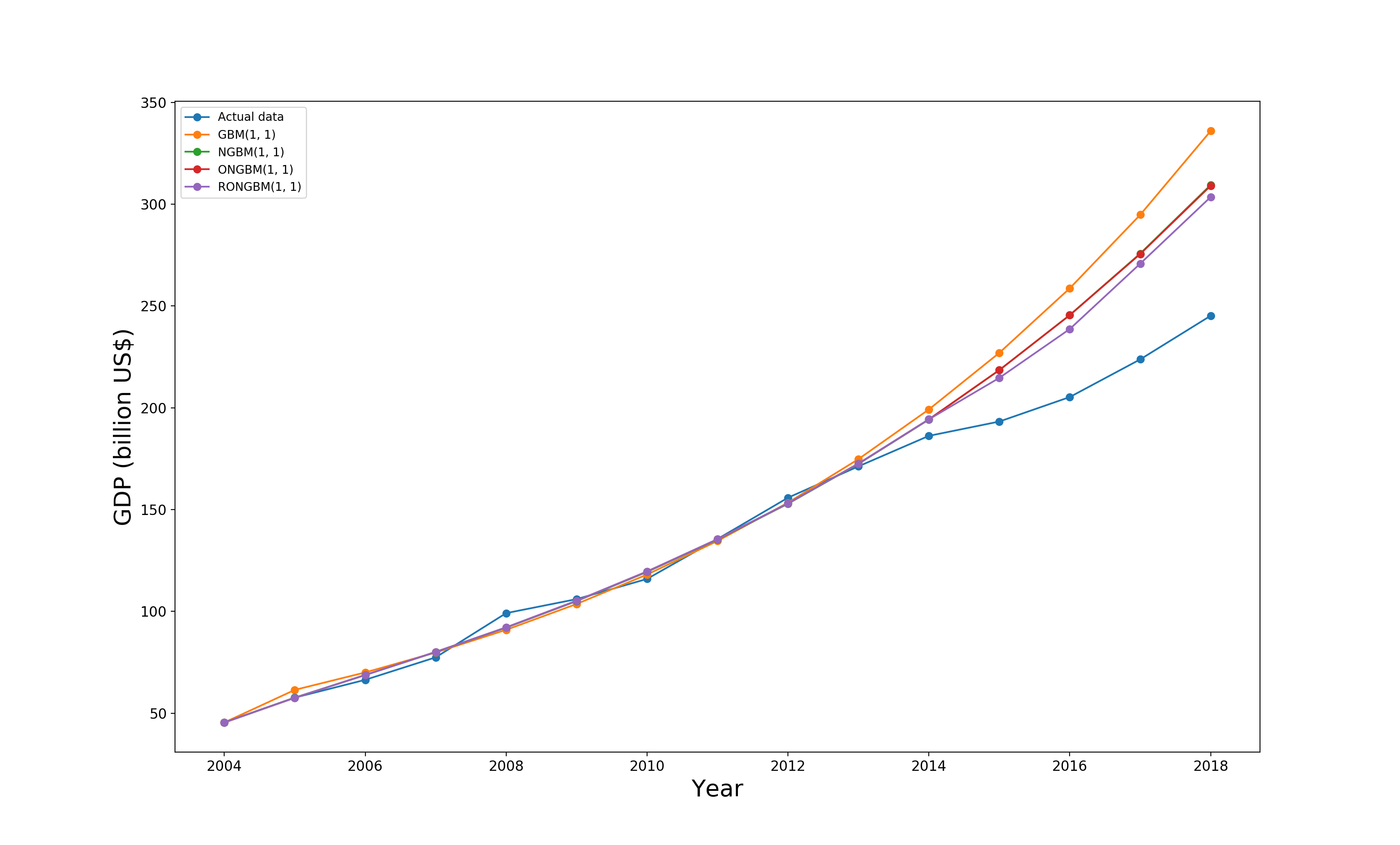}
    \label{fig:fig4}
\end{figure}

\subsection{Applying and benchmarking RONGBM(1, 1) in predicting 2019 - nCoV total infected cases}

\subsubsection{Context}
Right before the beginning of 2020, a new strain of novel coronavirus (COVID-19) was found from  of patients in the area of Wuhan City, Hubei Province, China. Full-length genome sequences were obtained from patients at early state showed that this new type of disease shared 79.5\% sequence identify to SARS-CoV in 2003. This COVID-19 genome could also be traced back to bats coronavirus \cite{read2020nCoV,zhou2020nCoV,li2020nCoV}. Researchers ared scared that this could be a new epidemic similar to the SARS-CoV and MERS-CoV epidemic in 2003 and 2012 separately that terrified worldwide. 
    \begin{figure}
        \centering
        \caption{Comparison between COVID-19, SARS-CoV, MERS-CoV and other outbreaks \cite{munster2020nCoV}}
        \includegraphics[scale=0.3]{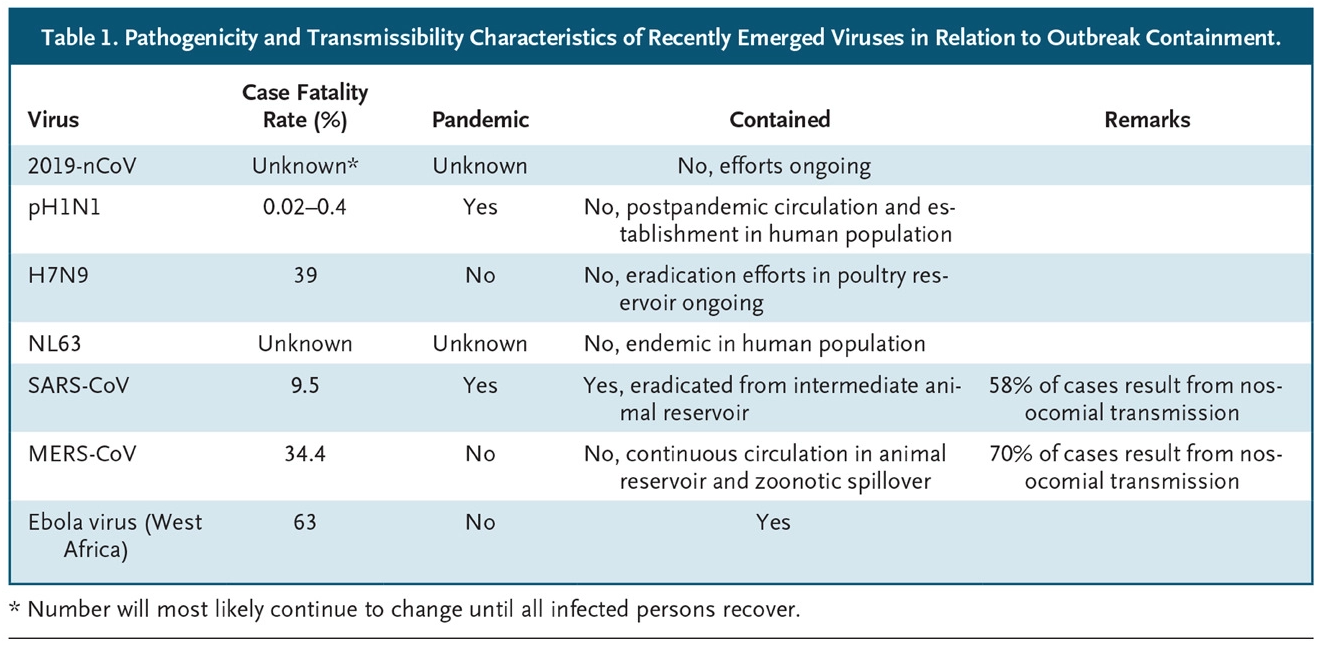}
        \label{fig:fig5}
    \end{figure}
    
\subsubsection{Methodology}
The total infected cases were retrieved free of charge from Worldometer (\url{https://www.worldometers.info/coronavirus/coronavirus-cases/}). We decided to use \texttt{2020-01-28} as the starting data point, as it would be more precise in terms of medical data collecting and analysis. 

Apart from comparing the Grey models, alternative modern machine learning models are also taken into account to compare accuracy and efficiency with Grey models.

In this section, the Rolling Optimized NGBM(1, 1) will not be taken into account considering the ARPE. As the training data sample is relatively small, this model will only be used to predict from the available data, based on the significant efficiency it has proven comparing to the Optimized NGBM in the previous example.

Considering the press release given be Reuters \parencite{reuters2020peaknCoV} stating that the peak of the COVID-19 outbreak will happen at mid- or late-February, we will proceed to predict the total infected cases everyday globally, starting from \texttt{2020-02-09} to \texttt{2020-02-18}.

\subsubsection{Results and analysis}

Considering 5 different models predicting within the time range \texttt{2020-01-28} to \texttt{2020-02-08}, the Optimized NGBM(1, 1) and even the traditional NGBM(1, 1) perform significantly better than two modern machine learning model (ANN and LSTM). This is due to the fact illustrated previously that NGBM(1, 1) yields relatively high accuracy upon handling short one - dimensional time series data. 

Within the next $10$ days, from \texttt{2000-02-09}, similar to the example of predicting Vietnam's GDP $2013 - 2018$, the predictions of the rolling optimized NGBM(1, 1) are lower than that of ONGBM(1, 1). This is due to the update of new data, reducing the increasing rate. However, both of the two predictions confirm that the peak of this COVID-19 outbreak will still not be happening within the next $10$ days. From plot ~\ref{fig:fig9}, it can be easily derived that the new infected cases by day is increasing, and there has been no signs of stopping. 
\begin{table}
    \begin{center}
    
    \caption{Results of ANN and LSTM in predicting COVID-19 infected cases by day}
    \label{table:2019nCoVANNLSTM}
    \begin{tabular}{|c|c|cc|cc|}
    \hline
\multirow{4}{4em}{\centering Date} & \multirow{4}{6em}{\centering Actual Data} & & & & \\ 
 &  & \multicolumn{2}{c|}{ANN} & \multicolumn{2}{c|}{LSTM} \\ 
    & & & & & \\
 &  & \multirow{2}{5em}{Prediction} & RPE & \multirow{2}{5em}{Prediction} & RPE \\ 
 &  &  & (\%) &  & (\%) \\ 
 
 \hline\hline

 \texttt{2020-01-28} & $6061$ & $6061$ & $0.00$ & $6061$ & $0.00$ \\ 
 \texttt{2020-01-29} & $7816$ & $8836$ & $13.05$ & $8707$ & $11.40$ \\
 \texttt{2020-01-30} & $9821$ & $10184$ & $3.70$ & $10213$ & $4.00$ \\
 \texttt{2020-01-31} & $11948$ & $12395$ & $3.74$ & $12486$ & $4.50$ \\
 \texttt{2020-02-01} & $14551$ & $14481$ & $-0.48$ & $14621$ & $0.48$ \\
 \texttt{2020-02-02} & $17387$ & $17371$ & $0.09$ & $17464$ & $0.44$ \\
 \texttt{2020-02-03} & $20626$ & $21294$ & $3.23$ & $21269$ & $3.11$ \\
 \texttt{2020-02-04} & $24553$ & $23061$ & $-6.07$ & $22928$ & $-6.62$ \\
 \texttt{2020-02-05} & $28276$ & $27808$ & $-1.65$ & $27837$ & $-1.55$ \\
 \texttt{2020-02-06} & $31439$ & $31649$ & $0.67$ & $31717$ & $0.88$ \\
 \texttt{2020-02-07} & $34875$ & $34727$ & $-0.43$ & $34757$ & $-0.34$ \\
 \texttt{2020-02-08} & $37552$ & $38285$ & $1.95$ & $38197$ & $1.72$ \\

\hline\hline

ARPE &  &  & $\textbf{2.92}$ &  & $\textbf{2.92}$ \\
\hline
\end{tabular} 
\end{center}
       
\end{table}

\begin{figure}
    \centering
    \caption{Total COVID-19 infected cases prediction by ANN and LSTM (from \texttt{2020-01-28} to \texttt{2020-02-08})}
    \includegraphics[scale=0.3]{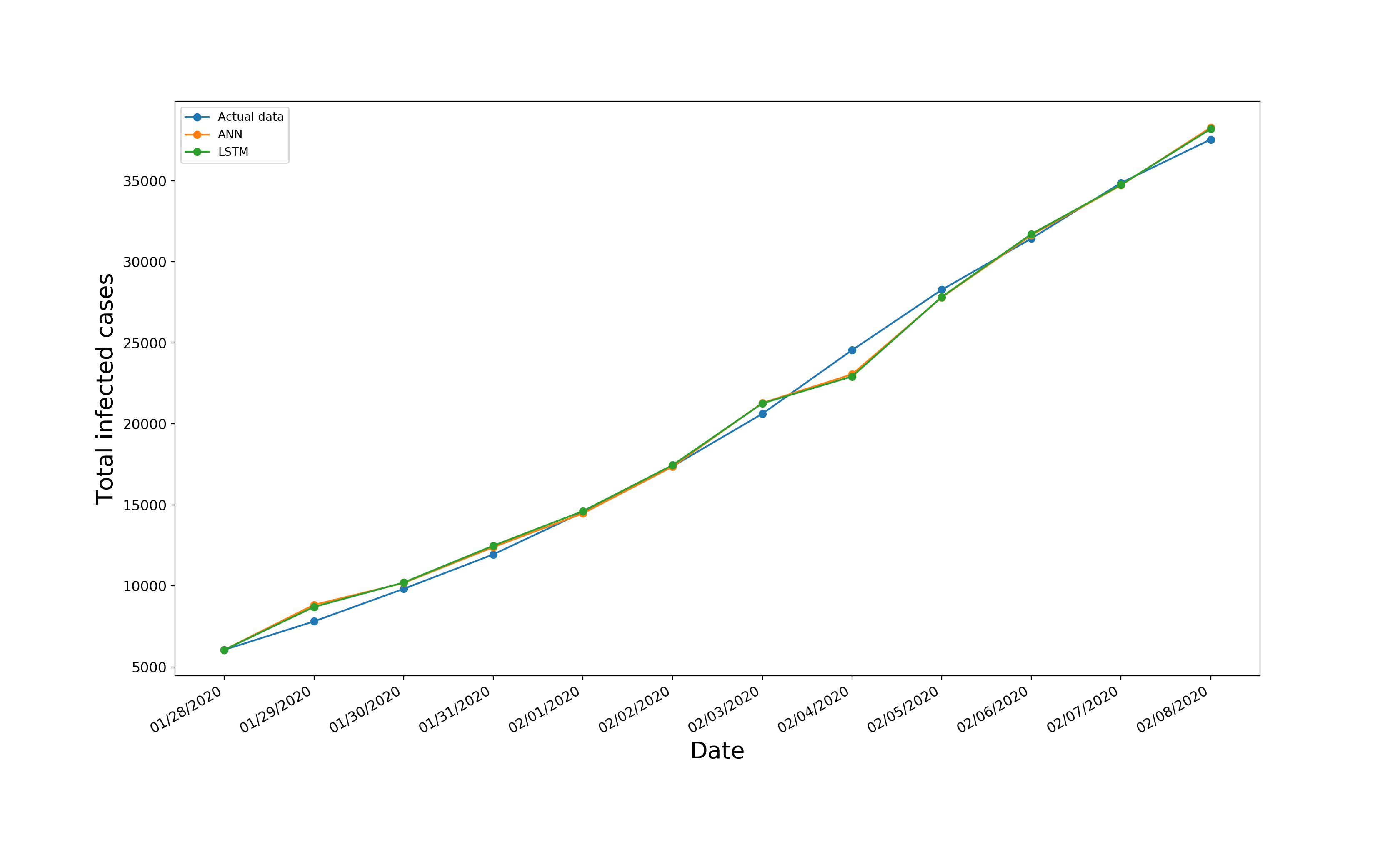}
    \label{fig:fig6}
\end{figure}

\begin{table}
    \begin{center}
    
    \caption{Results of GM(1, 1), NBGM(1, 1) and ONGBM(1, 1) in predicting COVID-19 infected cases by day}
    
    \begin{tabular}{|c|c|cc|cc|cc|}
    \hline
\multirow{5}{4em}{\centering Year} & \multirow{5}{6em}{\centering Actual Data} &  &  & \multicolumn{2}{c}{\textbf{ }}& \multicolumn{2}{|c|}{Optimized} \\ 

 &  &  &  & \multicolumn{2}{c}{NGBM(1, 1)} & \multicolumn{2}{|c|}{NGBM(1, 1)} \\ 

 &  &  \multicolumn{2}{c}{GM(1, 1)} & \multicolumn{2}{|c}{$n=0.41$} & \multicolumn{2}{|c|}{$n=0.505,P=0.7$} \\ 
 &  &  &  &  &  &  & \\ 
 &  & \multirow{2}{5em}{Prediction} & RPE & \multirow{2}{5em}{Prediction} & RPE & \multirow{2}{5em}{Prediction} & RPE \\

 &  &  & (\%) &  & (\%) &  & (\%) \\ 
 
 \hline\hline

 \texttt{2020-01-28} & $6061$ & $6061$ & $0.00$ & $6061$ & $0.00$ & $6061$ & $0.00$\\ 
 \texttt{2020-01-29} & $7816$ & $9946$ & $27.25$ & $7258$ & $-7.14$ & $7130$ & $-5.44$\\
 \texttt{2020-01-30} & $9821$ & $11451$ & $16.60$ & $9822$ & $0.01$ & $9824$ & $0.03$ \\
 \texttt{2020-01-31} & $11948$ & $13185$ & $10.35$ & $12418$ & $3.93$ & $12378$ & $3.60$ \\
 \texttt{2020-02-01} & $14551$ & $15181$ & $4.33$ & $15098$ & $3.76$ & $15056$ & $3.47$ \\
 \texttt{2020-02-02} & $17387$ & $17479$ & $0.53$ & $17898$ & $2.93$ & $17860$ & $2.72$ \\
 \texttt{2020-02-03} & $20626$ & $20125$ & $-2.43$ & $20842$ & $1.05$ & $20793$ & $0.81$ \\
 \texttt{2020-02-04} & $24553$ & $23172$ & $-5.62$ & $23953$ & $-2.44$ & $23862$ & $-2.82$ \\
 \texttt{2020-02-05} & $28276$ & $26679$ & $-5.64$ & $27251$ & $-3.62$ & $27068$ & $-4.27$ \\
 \texttt{2020-02-06} & $31439$ & $30719$ & $-2.29$ & $30755$ & $-2.17$ & $30417$ & $-3.25$ \\
 \texttt{2020-02-07} & $34875$ & $35369$ & $1.42$ & $34483$ & $-1.12$ & $33915$ & $-2.75$ \\
 \texttt{2020-02-08} & $37552$ & $40724$ & $8.45$ & $38455$ & $2.40$ & $37564$ & $0.03$ \\

\hline\hline

ARPE &  &  & $\textbf{7.08}$ &  & $\textbf{2.55}$ &  & $\textbf{2.43}$ \\
\hline
\end{tabular} 
\end{center}
\label{table:2019nCoV3models}
\end{table}

\begin{figure}[h]
    \centering
    \caption{Total COVID-19 infected cases predicted by GM(1, 1), NGBM(1, 1) and ONGBM(1, 1) (\texttt{2020-01-28} - \texttt{2020-02-08})}
    \includegraphics[scale=0.3]{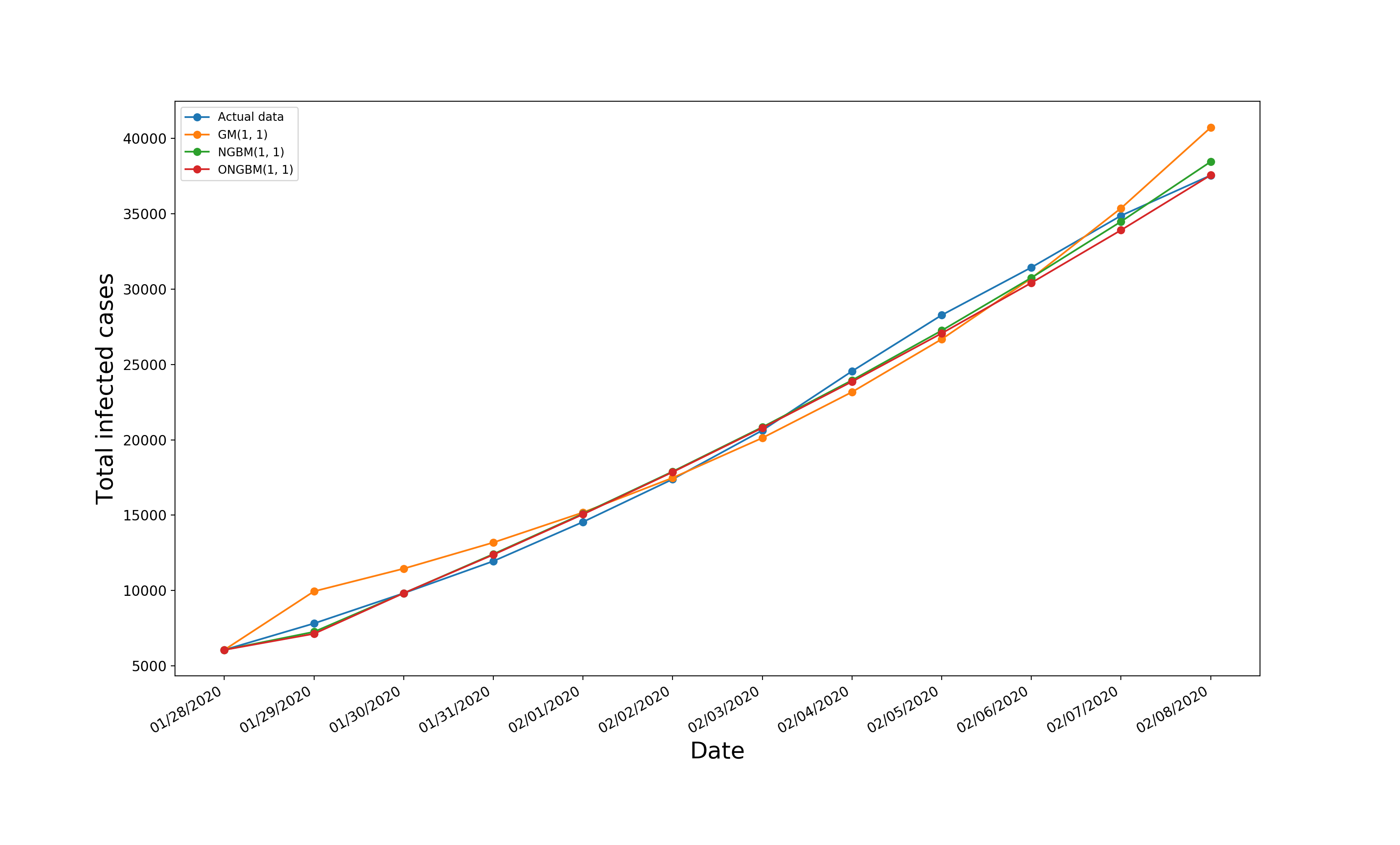}
    \label{fig:fig7}
\end{figure}

\begin{table}
    \begin{center}
    \caption{Predictions by ONGBM(1, 1) and RONGBM(1, 1) of COVID-19 infected case by day}
    \begin{tabular}{|c|c|cc|c|}
    \hline
\multirow{4}{4em}{\centering Year} & & \multicolumn{3}{c|}{Rolling Optimized} \\ 
 &  Optimized NGBM(1, 1) & \multicolumn{3}{c|}{NGBM(1, 1)} \\ 
 &  $n = 0.505, P = 0.7$ &  \multirow{2}{5em}{Parameter $P$} & \multirow{2}{5em}{Parameter $n$} & \multirow{2}{5em}{Prediction} \\ 
 &  &  &  & \\ 
 
 \hline\hline
 \texttt{2000-02-09} & $41373$ & $0.7$ & $0.505$ & $41373$ \\
 \texttt{2000-02-10} & $45344$ & $0.55$ & $0.47$ & $45050$ \\
 \texttt{2000-02-11} & $49484$ & $0.505$ & $0.36$ & $48461$ \\
 \texttt{2000-02-12} & $53799$ & $0.5$ & $0.285$ & $52176$ \\
 \texttt{2000-02-13} & $58295$ & $0.49$ & $0.22$ & $56045$ \\
 \texttt{2000-02-14} & $62978$ & $0.47$ & $0.14$ & $60493$ \\
 \texttt{2000-02-15} & $67854$ & $0.49$ & $0.095$ & $65041$ \\
 \texttt{2000-02-16} & $72930$ & $0.485$ & $0.065$ & $70001$ \\
 \texttt{2000-02-17} & $78214$ & $0.49$ & $0.045$ & $75238$ \\
 \texttt{2000-02-18} & $83711$ & $0.49$ & $0.035$ & $80848$ \\

\hline
\end{tabular} 
\end{center}
\label{table:2019nCoVpredict10daysRONGBM}
\end{table}

\begin{figure}
    \centering
    \caption{Total COVID-19 infected cases predicted by ONGBM(1, 1) and RONGBM(1, 1) (\texttt{2020-01-28} to \texttt{2020-02-18})}
    \includegraphics[scale=0.3]{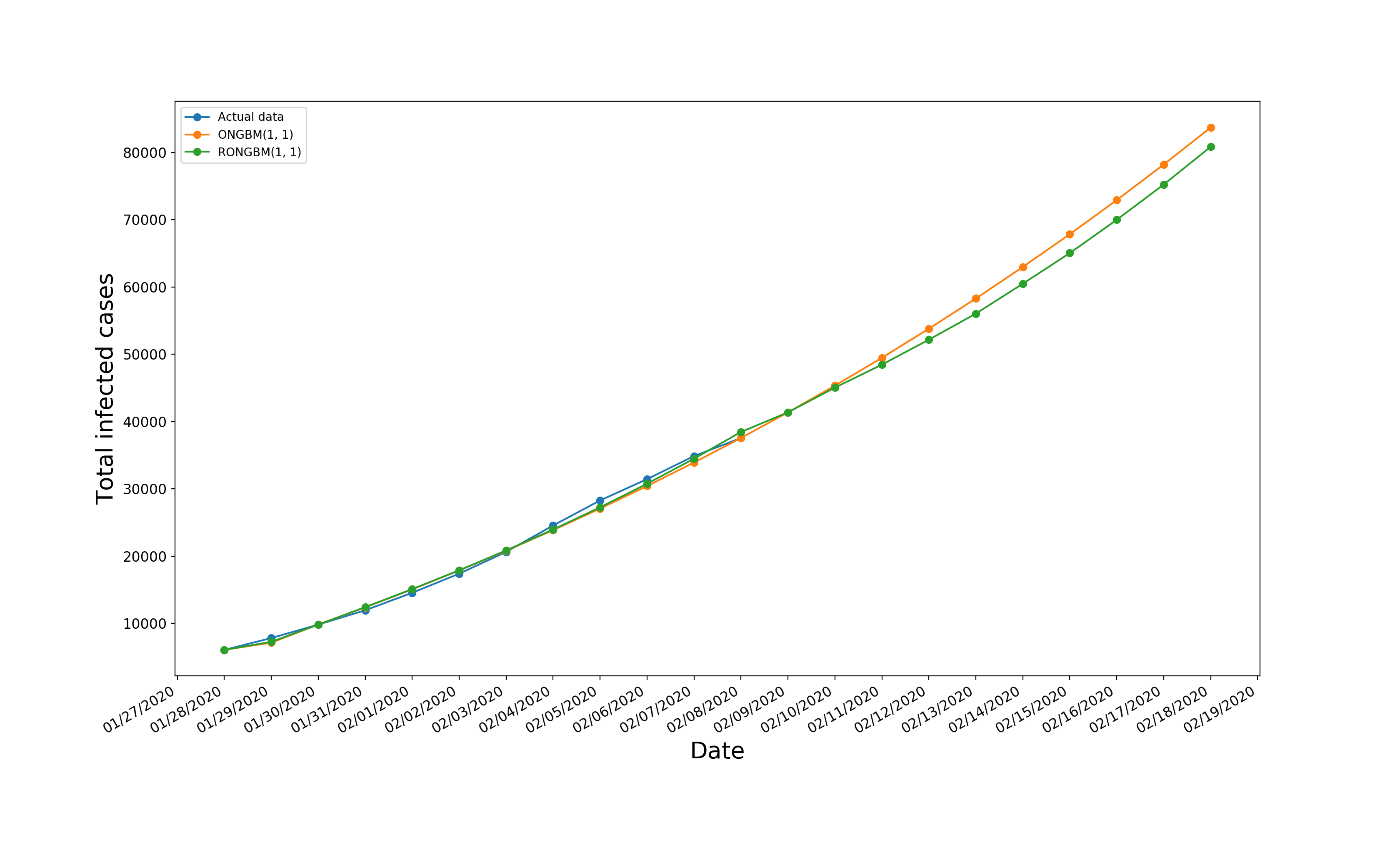}
    \label{fig:fig8}
\end{figure}

\begin{figure}
    \centering
    \caption{COVID-19 newly infected cases by day predicted by ONGBM(1, 1) and RONGBM(1, 1) (\texttt{2020-02-09} to \texttt{2020-02-18})}
    \includegraphics[scale=0.3]{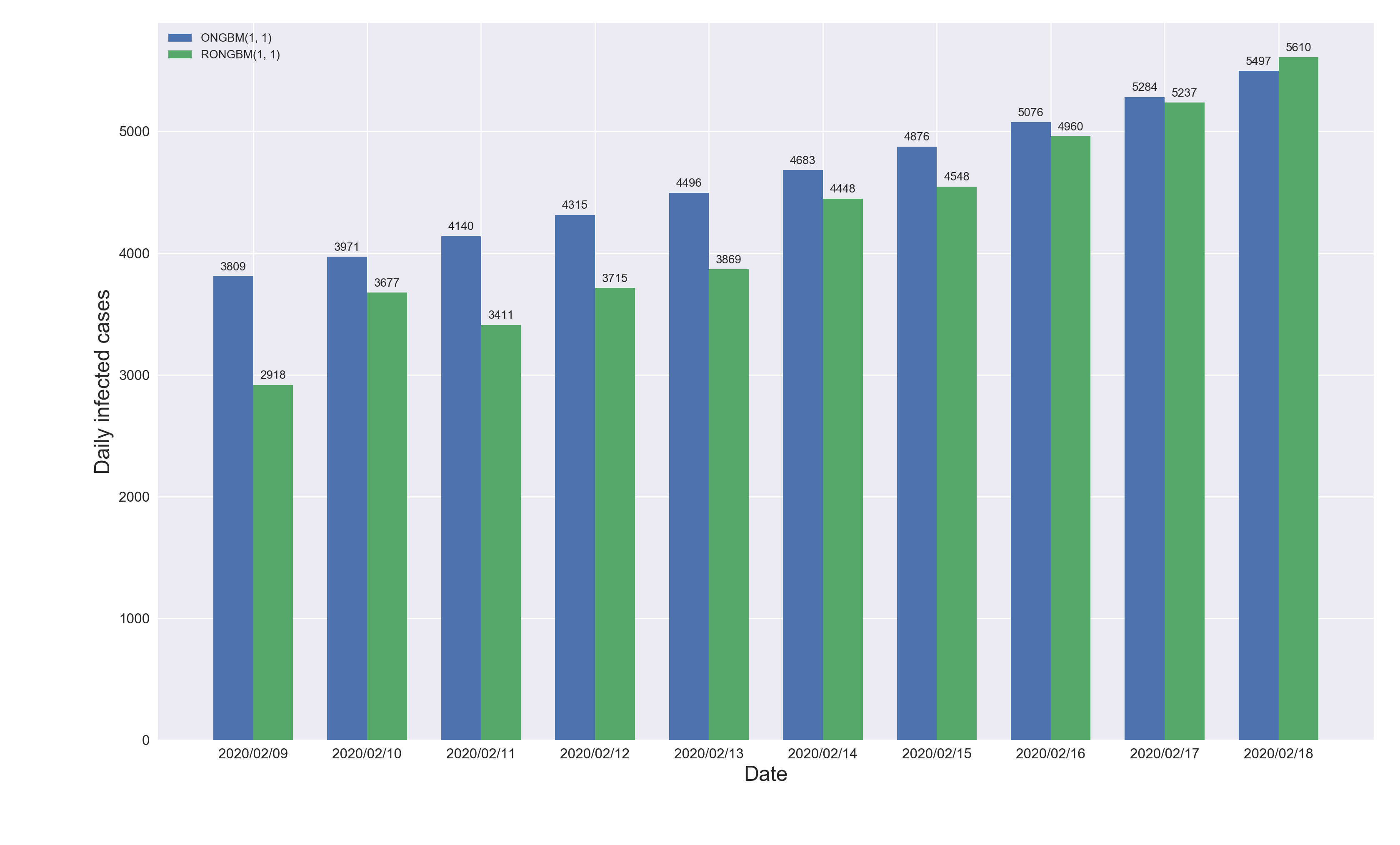}
    \label{fig:fig9}
\end{figure}

\section{Conclusion and future suggestions}
In this paper, a new rolling optimized NGBM(1, 1) model has been proposed. There are two following major advantages of this model in comparison with the existing ones:
    \begin{itemize}
        \item Within the proposed model, all of the parameters are optimized simultaneously, including the exponential parameter, the background value and the initial condition.
        \item The first application of the rolling mechanism upon the combination of all currently available methods of parameter optimization.
    \end{itemize}

The efficiency of this newly proposed model has also been tested with the other Grey models (GM(1, 1), NGBM(1, 1) and ONGBM(1, 1)), which showed significant improvement. Later, while this model is applied to predict the total COVID-19 infected cases, it has also been put into comparison with the latest and most effective machine learning models, including Artificial Neural Network (ANN) and Long Short-Term Memory (LSTM). Within the context of extremely limit amount of data and time constrain, RONGBM(1, 1) outperforms both of the two other models, which additionally confirms the effectiveness of Grey models upon dealing with small amount of data.

From the results obtained by applying RONGBM(1, 1) on available data until the final day of retrieval (\texttt{2020-02-10}), there are now signs that the outbreak of COVID-19 will reach its peak within the next $10$ days as expected. The number of newly infected patients reported everyday will even be higher than that of the present, increasing day by day. However, as Grey models are only applied on one-dimensional time series data, it is impossible to detect or predict changes due to any other related or outside effects (medical advancement, natural conditions, etc.). As a result, with enough amount of background knowledge into the field, applying a multivariate Grey model would yield more accurate and medically meaningful results.

Considering the proposed model RONGBM(1, 1) itself, there are two major problems that should be discussed and investigated in the future:
    \begin{itemize}
        \item \textbf{Training data size:} As the training data size has a significant impact on any Grey model \parencite{wu2013samplesize}, the amount of data used in the COVID-19 infected case estimation has been manually chosen. In the future, for Grey models to reach maximum optimality, there should be a formal, mathematically derived procedure to define the optimal size of training data. 
        \item \textbf{Parameter optimization:} Within the proposed model, in order to obtain the optimization of the two parameters $P$ and $n$, two loops have been run through. Due to the resource constrain of the current computational tools, the step size of these loops are still large enough to omit actual optimal values, leading to the model not perfectly optimized. Moreover, using the proposed formula, as demonstrated within the practical section, does not yield the optimal error. As a result, in the future, there should be a mathematically exact way to derive the parameter values optimally.
    \end{itemize}

\section*{Conflicts of Interest}
The authors declare that there are no conflicts of interest regarding the publication of this paper.

%\section*{Acknowledgement}
%The authors would like to thank Professor Mehmet Dik, Visiting Professor at Department of Mathematics and Computer Science, Beloit College, for providing initial ideas and support during the process of conducting this research.

%\bibliographystyle{unsrt}  


\printbibliography

\end{document}